\documentclass[12pt, a4paper]{iopart}
\usepackage{graphicx}
\usepackage{subfig}
\usepackage{iopams}
\usepackage{setstack}
\usepackage{cite}
\usepackage{psfrag}
\usepackage{wasysym}
\newcommand*{\pbar}{\ensuremath{\overline{\rm{p}}}}
\newcommand*{\pbHe}{\ensuremath{\overline{\rm{p}}\rm{He}^+}}

\begin{document}

\title{Improved Study of the Antiprotonic Helium Hyperfine Structure}
\author{T Pask$^{1}$, D Barna$^{2,3}$, A Dax$^{2}$, R S Hayano$^2$, M Hori$^{2}$\footnote{Present address: Max-Planck-Institut f\"{u}r Quantenoptik, Hans-Kopfermann-Strasse 1, D-85748 Garching, Germany.}, D Horv\'ath$^{3, 4}$, B Juh\'asz$^1$, C Malbrunot$^1$\footnote{Present address: TRIUMF, 4004 Wesbrook Mall, Vancouver, BC, V6T 2A3, Canada.}, J Marton$^{1}$, N Ono$^2$, K Suzuki$^1$, J Zmeskal$^1$ and E Widmann$^1$}

\address{$^1$ Stefan Meyer Institute for Subatomic Physics, Austrian Academy of Sciences, Boltzmanngasse 3, A-1090 Vienna, Austria.} \ead{thomas.pask@cern.ch}

\address{$^2$ Department of Physics, University of Tokyo, 7-3-1 Hongo, Bunkyo-ku, Tokyo 113-0033, Japan}

\address{$^3$ KFKI Research Institute for Particle and Nuclear Physics, H-1525 Budapest, PO Box 49, Hungary}

\address{$^4$ Institute of Nuclear Research of the Hungarian Academy of Sciences, H-4001 Debrecen, PO Box 51, Hungary}

\begin{abstract}

We report the initial results from a systematic study of the hyperfine (HF) structure of antiprotonic helium $(n,l)$~=~(37,~35) carried out at the Antiproton Decelerator (AD) at CERN.  We performed a laser-microwave-laser resonance spectroscopy using a continuous wave (cw) pulse-amplified laser system and microwave cavity to measure the HF transition frequencies.  Improvements in the spectral linewidth and stability of our laser system have increased the precision of these measurements by a factor of five and reduced the line width by a factor of 3 compared to our previous results.  A comparison of the experimentally measured transition frequencies with three body QED calculations can be used to determine the antiproton spin magnetic moment, leading towards a test of CPT invariance.

\end{abstract}

\submitto{\jpb}

\maketitle

\section{Introduction}

It was first discovered at KEK, in 1991~\cite{Iwasaki}, and later measured at the Low Energy Antiproton Ring (LEAR)~\cite{Yam93, blue} and the Antiproton Decelerator (AD) at CERN~\cite{newblue}, that a small fraction ($\sim$~3\%) of antiprotons ($\pbar$), stopped in a helium target, survive to form antiprotonic helium ($\pbHe$): an exotic metastable atom consisting of a helium nucleus, an electron and an antiproton (He$^{++}$ + e$^-$ + $\pbar$)~\cite{Yam92, condo, russell}.  The antiprotons occupy highly excited states in which they are temporarily protected from annihilation.  They undergo a cascade by which the $\pbar$ radiatively de-excites to a lower-lying state, with a lifetime of the order of microseconds.  A \emph{hyperfine structure (HFS)}~\cite{blue} arises from the coupling of the electron and the antiproton magnetic moments.  This paper presents the initial results of a current systematic study where the method of measuring the HFS of $\pbHe$ has been improved.  The new results reduce the error compared to the previous measurements and converge towards current theories.

Previously, laser spectroscopy has been performed on the various levels of the cascade~\cite{blue,Torii99,Hori01,Hori03,pe}. Two types of electric dipole transitions have been studied: favoured $(\Delta v = 0; (n,l) \rightarrow (n \, - \, 1, \, l \, - \, 1))$ and  unfavoured $(\Delta v \, = \, 2; \, (n,l) \, \rightarrow \, (n \, + \, 1,\, l \, - \, 1))$~\cite{blue} where $n$, $l$ and $v \, = \, n \, - \, l \, - \, 1$ are the principal, total angular momentum and vibrational quantum numbers, respectively.  The dipole moment of the unfavoured transitions is an order of magnitude smaller than that of the favoured.  The unfavoured ones have a hyperfine splitting $\Delta_{\rm{HF}}$~=~1.5~-~1.8~GHz, while the favoured $\Delta_{\rm{HF}}$~$\leq$~0.5~GHz.  Improvements to the laser system have reduced the laser linewidth from 800~MHz to 100~MHz but the resolution remains limited by Doppler broadening (300~-~500~MHz at 6~K).  From among the candidate transitions, the $(37, \, 35) \, \rightarrow \, (38,~34)$ unfavoured transition was chosen for study because the parent state is relatively highly populated, containing some 0.3\% of the antiprotons stopped in the target \cite{Hori02}.

In 1996, at LEAR, a precise laser spectroscopy scan at 726.10~nm was made, resolving, for the first time, the two hyperfine peaks of the $(37, \, 35) \, \rightarrow \, (38,~34)$ transition~\cite{Wid97}.  However, to accurately determine the HF splitting it is necessary to induce a transition between the HFS substates.  While it is possible  to induce a transition between $(n,l)$ states with laser light, a magnetic M1 resonance is required to move between the HFS substates.

In 2001, a laser-microwave-laser measurement was performed, resolving the hyperfine structure of the (37,~35) state to a precision of 300~kHz~\cite{HFS}.  By comparing the experimentally measured transition frequency with three body QED calculations, the antiproton spin magnetic moment can be extracted.  Because this is small compared to the other contributions, the precision of the HFS measurement does not directly convert to the same precision in the spin magnetic moment.

A break down of this article is as follows: Section~\ref{sec:HFS} describes the HFS and the transitions that can be induced, Sections~\ref{sec:Method} and~\ref{sec:App} detail the experimental method and apparatus respectively, in Section~\ref{sec:Results} the results are presented and discussed, while Section~\ref{sec:Con} contains our conclusions.

\section{The hyperfine structure of antiprotonic helium}\label{sec:HFS}

The HFS of ${\overline{\rm{p}}^4\rm{He}^+}$ arises from the interactions of the antiproton orbital angular momentum $\vec{L}$, the electron spin $\vec{S}_{\rm{e}}$ and the antiproton spin $\vec{S}_{\pbar}$.  It has been first calculated by Korobov and Bakalov~\cite{KB} who constructed an effective Hamiltonian

\begin{displaymath} \fl
H^{\rm{eff}} = E_{1}(\vec{L} \cdot \vec{S}_{\rm{e}}) + E_{2}(\vec{L} \cdot \vec{S}_{\pbar}) + E_{3}(\vec{S}_{\rm{e}} \cdot \vec{S}_{\pbar}) 
\end{displaymath}

\begin{equation} + E_{4} \{ 2L(L + 1)(\vec{S}_{\rm{e}} \cdot \vec{S}_{\pbar}) - 6[ (\vec{L} \cdot \vec{S}_{\rm{e}})(\vec{L} \cdot \vec{S}_{\pbar})]\}.
\end{equation}

\begin{figure}[!h]
\begin{center}
\includegraphics[scale=0.6]{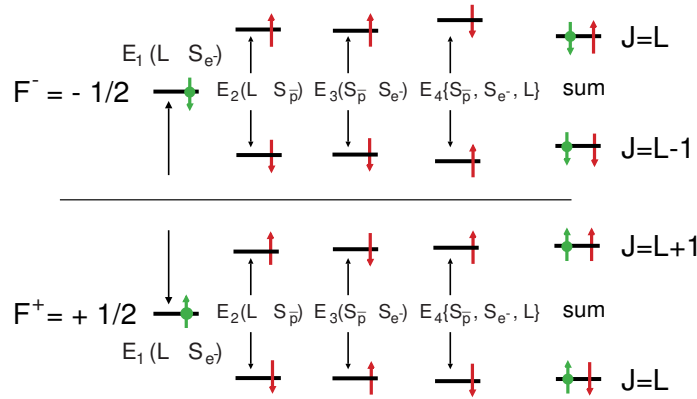}
\end{center}
\caption{Individual contributions to the SHF splitting, where $E_{1}(\vec{L} \cdot \vec{S}_{\rm{e}})$ is the $\pbar$~-~e$^{-}$ spin-orbit splitting, $E_{2}(\vec{L} \cdot \vec{S}_{\pbar})$ is the $\pbar$ spin-orbit splitting, $E_{3}(\vec{S}_{\pbar} \cdot \vec{S}_{\rm{e}})$ is the $\pbar$~-~e$^{-}$  scalar spin-spin splitting and $E_{4}\{\vec{S}_{\pbar}, \vec{S}_{\rm{e}}-\vec{L}\}$ is the $\pbar$~-~e$^{-}$ tensor spin-spin splitting.  The circled and straight arrows represent the electron and antiproton spin directions respectively.}
\label{fig:hierarchy}
\end{figure}

\begin{figure}[!h]
\begin{center}
\includegraphics[scale=0.6]{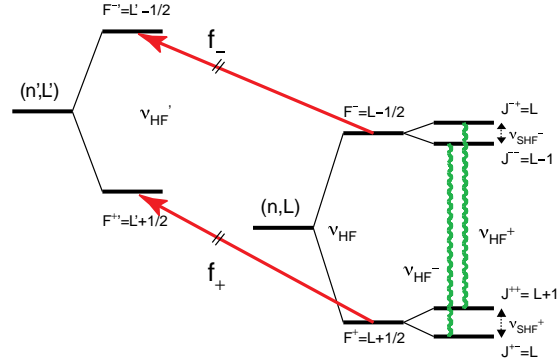}
\end{center}

\caption{Schematic view of the level splitting of $\pbHe$ for the unfavoured electric dipole transitions.  The state drawn on the right is the radiative decay dominated parent $(n,L)$, and the left state is the Auger decay dominated daughter $(n',L')$. The laser transitions, from the parent to daughter doublets, are indicated by straight lines and the microwave transitions, between the quadruplets of the parent, by wavy ones.  For this experiment $(n,L)$~=~(37,~35) and $(n',L')$~=~(38,~34).}
\label{fig:trans}
\end{figure}

Due to the large orbital angular momentum of the antiproton $(L$~$\sim$~35), the dominant HF splitting arises from the interaction of orbital angular momentum with the electron spin (first term: $\pbar$~-~e$^{-}$ spin-orbit splitting). The contributions of the following three terms cause a \emph{superhyperfine (SHF)} splitting~\cite{blue} of each level of the HF doublet.  The contact spin-spin (third term) and the tensor spin-spin (fourth term) interactions act to almost cancel each other out, leaving the coupling of the orbital angular momentum and the antiproton spin (second term: $\pbar$ spin-orbit splitting) as the dominant contribution to the SHF splitting (figure~\ref{fig:hierarchy}).  The resulting quadruplet structure for $\pbHe$ is shown in figure~\ref{fig:trans}, where the HF splitting $\nu_{\rm{HF}} \, = \, 10-15$ GHz is two orders of magnitude larger than the SHF splitting ($\nu_{\rm{SHF}} \, = \, 150-300$ MHz) which is too small to be resolved by the current apparatus.

The HF doublet is described by the quantum number $\vec{F} \, = \, \vec{L} \, + \, \vec{S}_{e}$ with components $F_{+} \, = \, L \, + \, \frac{1}{2}$ and $F_{-} \, = \, L \, - \, \frac{1}{2}$.  The SHF quadruplet is described by $\vec{J} \, = \, \vec{F} \, + \, \vec{S}_{\pbar}$ with components $J_{-+} \, = \, F_{-} \, + \, \frac{1}{2}$, $J_{--} \, = \, F_{-} \, - \, \frac{1}{2}$, $J_{++} \, = \, F_{+} \, + \, \frac{1}{2}$ and $J_{+-} \, = \, F_{+} \, - \, \frac{1}{2}$.  Between these substates there are two M1 transitions $\nu^{+}_{\rm{HF}}$ and $\nu^{-}_{\rm{HF}}$ which cause an electron spin flip and can be induced by an oscillating magnetic field:

\numparts
\begin{equation}
\nu^{+}_{\rm{HF}}: \, J_{++} \, = \, F_{+} \, + \, \frac{1}{2} \, = \, L \, + \, 1 \, \leftrightarrow \, J_{-+} \, = \, F_{-} \, + \, \frac{1}{2} \, = \, L,
\end{equation}

\begin{equation}
\nu^{-}_{\rm{HF}}: \, J_{+-} \, = \, F_{+} \, - \, \frac{1}{2} \, = \, L \, \leftrightarrow \, J_{--} \, = \, F_{-} \, - \, \frac{1}{2} \, = \, L \, - \, 1.
\end{equation}
\endnumparts

\section{Laser-microwave-laser spectroscopy method}\label{sec:Method}

\begin{figure}[!h]
\begin{center}
\includegraphics[scale=0.4]{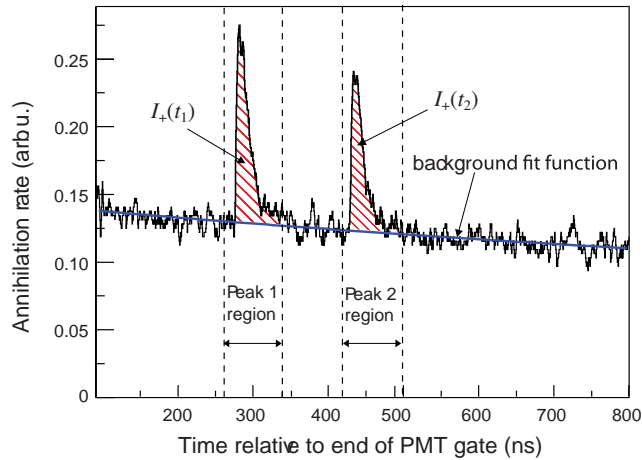}
\end{center}
\caption{Two laser stimulated annihilation peaks against the exponential decaying background of the other state populations.  The peak-to-total of each is calculated by taking the ratio of the peak area ($I_+$) to the total area under the full spectrum.}
\label{fig:adats}
\end{figure}

A narrow-band laser pulse is tuned to the $f^{+}$ transition between the radiative decay dominated parent state (37,~35) and the Auger decay dominated daughter state (38,~34) shown in figure~\ref{fig:trans}.    The  daughter state is fast-decaying, with a lifetime (11~ns) \cite{Yam02} two orders of magnetude shorter  than that of the parent.  Therefore, when the laser is fired, it produces a sharp annihilation peak against the exponentially decaying background of the time spectrum (figure~\ref{fig:adats}).  The ratio of the peak area to this background (peak-to-total) indicates the size of the population transferred from the parent to the daughter state.

In addition to measuring the population, an asymmetry is produced such that the $F^{+}$ doublet is depopulated while the $F^{-}$ is unaffected.  The  microwave pulse follows which, if on-resonance, results in a partial inversion of the asymmetry and refills either the $J_{++}$ or $J_{+-}$ from the $J_{-+}$ or $J_{--}$ state respectively.  A second laser pulse, tuned to the same $f^{+}$ transition, depopulates the doublet again, producing a second annihilation peak in the time spectrum (figure~\ref{fig:adats}).

The first laser-induced annihilation peak remains constant (figure~\ref{fig:adats}), fluctuating only statistically or with the varying conditions of the target, antiproton beam, or laser pulse, and is therefore used to normalise the second.  The second annihilation peak has a maximum as the microwave is scanned through resonance and corresponds to the population transferred between the hyperfine substates.

\section{Experimental apparatus}\label{sec:App}

The experiment was carried out at the AD at CERN, which delivered a pulse of $1-4 \, \times \, 10^7$ antiprotons with a length of 200~ns (FWHM) and kinetic energy 5.3~MeV every $\sim$~90~s.  The setup is described in detail in Sakaguchi \emph{et al.}~\cite{App}.  At extraction the antiprotons were stopped in a gas target at a temperature of 6.1~K and a pressure of 250~mbar (number density 3~$\times$~$10^{20}$~cm$^{-3}$).  During the experiment the profile of the $\pbar$ beam was monitored by a nondestructive beam profile monitor (BPM)~\cite{BPM} positioned $\sim$~20~cm upstream of the target.

The charged pions, produced by the antiprotons annihilating in the helium nucleus, were detected by two Cherenkov counters covering 1.5$\pi$ steradians around the target.  The signal was amplified and detected by fine-mesh photomultipliers (PMTs) (Hamamatsu Photonics R5505 GX-ASSY2).  The resulting analog delayed annihilation time spectrum (ADATS) was recorded in a digital oscilloscope (DSO). Since 97\% of the stopped antiprotons annihilate within nanoseconds, the PMTs were gated off during the $\pbar$ pulse arrival so that only the 3\% metastable tail was recorded~\cite{PMT}.

The pulse-amplified continuous wave (cw) laser system is a variation of the one used in ASACUSA's recent high precision laser spectroscopy measurement~\cite{pe}.  A cw laser beam of wave-length 726.1~nm is split into two seed beams.  The laser pulses were produced by amplifying the seeds using Bethune cells pumped by pulsed Nd:Yag lasers, the second delayed by time $T$ after the first.  The pump beams were stretched so that the two pulse lengths were of the same order as the Auger lifetime.

Advantages over the previous system include: (1)~a narrow linewidth ($\sim$~100~MHz), which allowed one HF doublet to be depopulated without affecting the other; (2)~single-mode and high shot-to-shot frequency stability ($\sim$~50-100~MHz) which reduced the quantity of statistics required for each measurement; (4)~a long  pulse length (18~ns for the first laser and 13~ns for the second laser) which increased the laser depopulation efficiency; and (5)~an arbitrary time difference between laser pulses, which was previously limited to 150~ns.  This determined the length of the microwave pulse and therefore the final line width which is dependent on the Fourier transform of the microwave pulse.  The emitted energy fluence at the target was $\sim$~30~mJ/cm$^2$ with a spot diameter of 5~mm resulting in a maximum depopulation efficiency of 80\% compared to the previous maximum of 50\%~\cite{Jun_PhD}.

\begin{figure}[!h]
\subfloat[]{
\label{fig:subfig:L2006}
\includegraphics[scale=0.50]{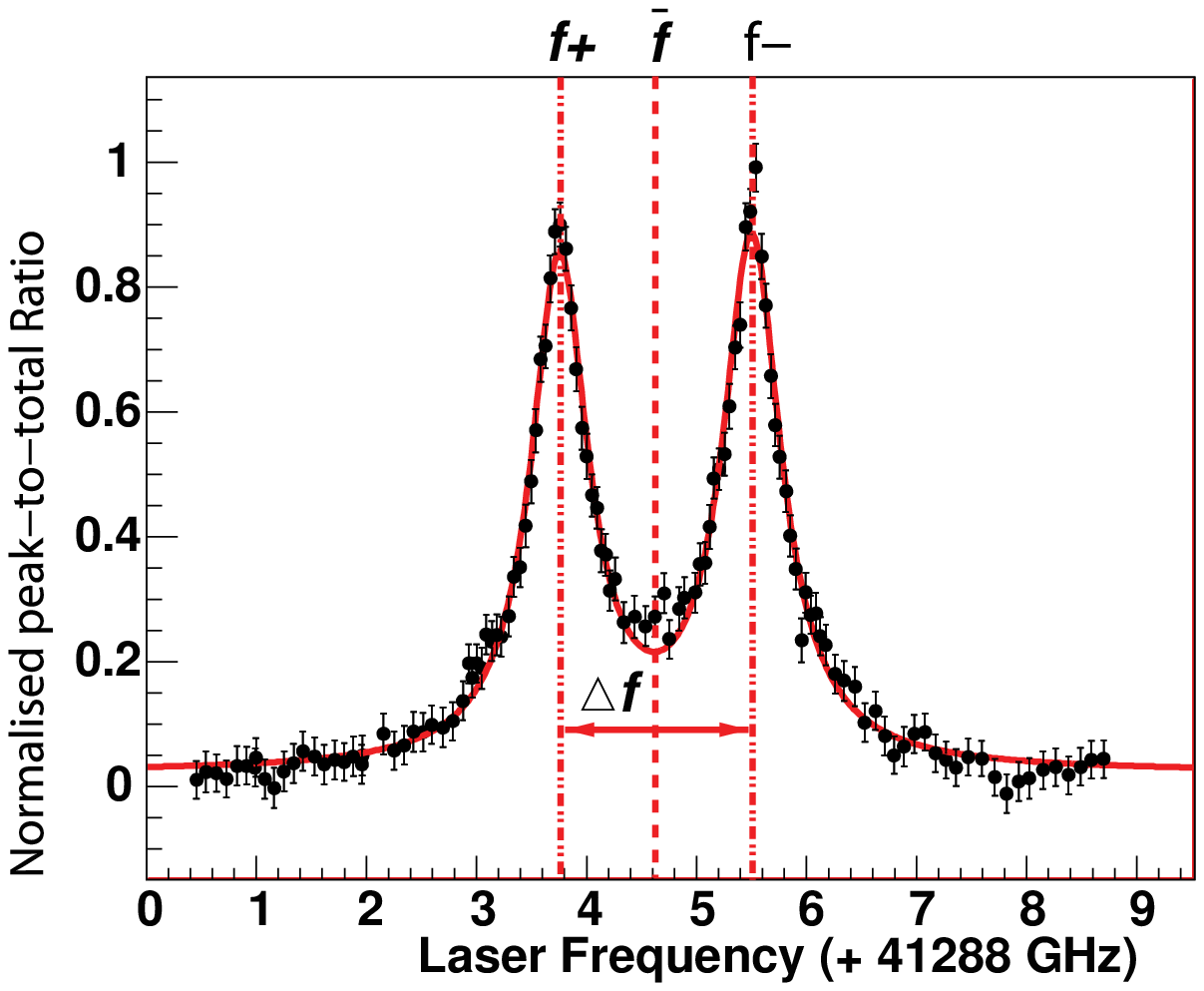}}
\subfloat[]{
\label{fig:subfig:L2001}
\includegraphics[scale=0.38]{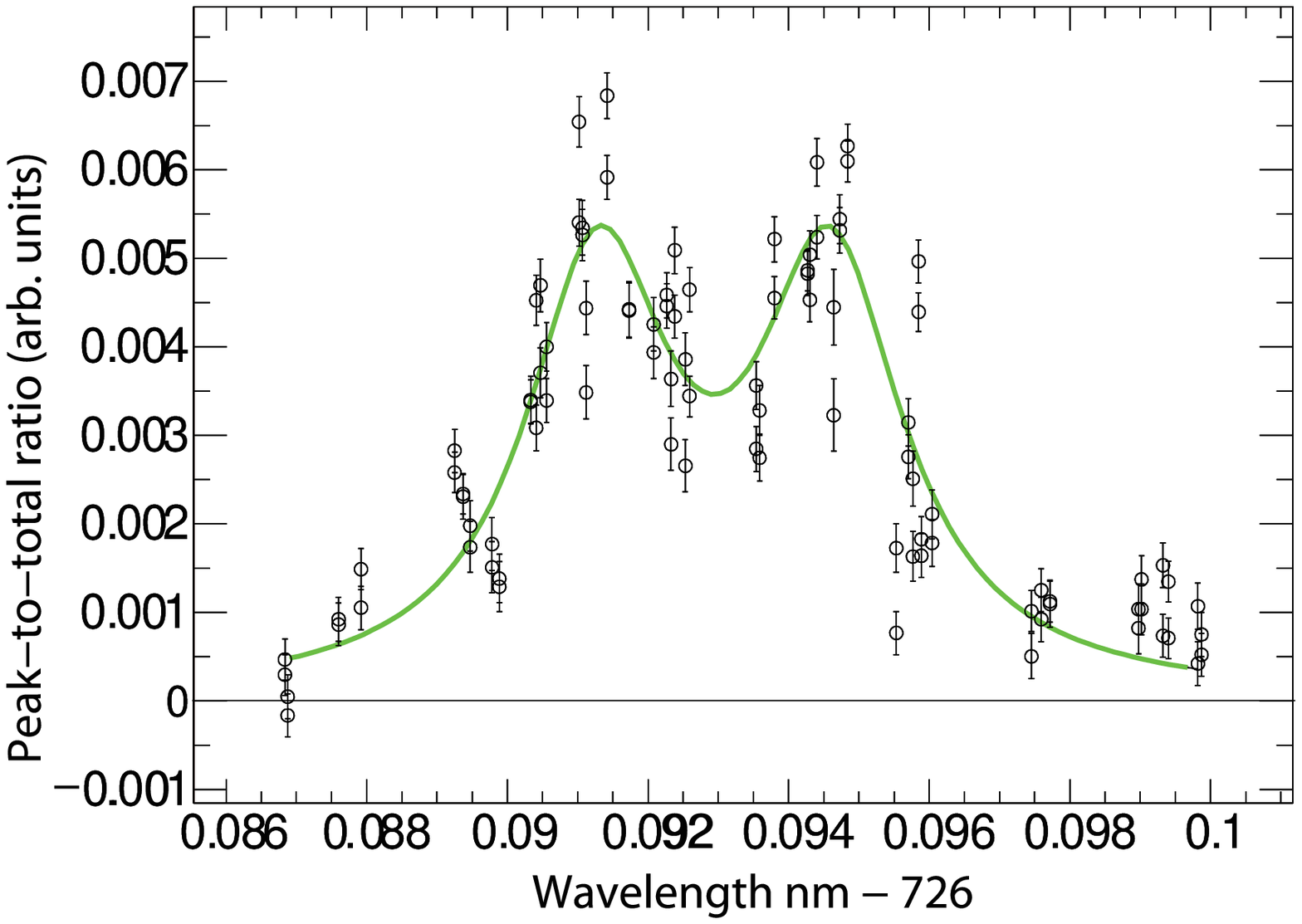}}

\caption{Laser resonance profile of the $(n,l)$~=~(37,~35) to (38,~34) transition measured at a temperature of 6.1~K, at He pressure of 250~mbar and laser energy fluence of 30~mJ/cm$^{2}$ in (a) 2006 fitted with a double Voigt profile and (b) 2001~\cite{HFS} fitted with a double Voigt profile.}
\label{fig:lscan}
\end{figure}

Figure~\ref{fig:lscan} shows an example of a laser scan taken with the recent laser setup (figure~\ref{fig:subfig:L2006}) compared with the old system (figure~\ref{fig:subfig:L2001}).  The improved stability is demonstrated by the fact that each data point represents a single shot from the AD (figure~\ref{fig:subfig:L2006}), whereas, previously, an average of several scans was required because the data points were more widely scattered (figure~\ref{fig:subfig:L2001}).  The band width is now small enough so that each peak can be individually resolved: a laser pulse tuned to the $f^+$ transition no longer partially depopulates the $F^-$ doublet.  The dominating effect was expected to be the Doppler broadening which has a Gaussian profile with a 320~MHz width.  However a Lorentzian profile was observed with a 350~MHz width.  It was concluded that each peak was probably narrowed through Dicke narrowing because the collision frequency was of the same order as the Rabi oscillations~\cite{Dicke, Dicke2}.

The current microwave apparatus is similar to that described in Sakaguchi \emph{et al.}~\cite{App}.  A short summary is provided here.  The microwave frequency was produced by an Anritsu 37225B vector network analyser (VNA) referenced to a 10~MHz GPS (HP~58503B) satellite signal which ensured a relative frequency uncertainty of $< \, 10^{-7}$.   The microwave signal was amplified by a pulsed  travelling wave tube amplifier (TWTA) (TMD PTC6358) and transported to the target by a rectangular waveguide.  The target was contained within a liquid helium-cooled cryostat and the section of waveguide that passed into the cryostat was constructed from stainless steel to reduce thermal conductivity.

A cylindrical resonant microwave cavity created the microwave field at the target.  To cover the entire microwave range, the cavity, which had an unloaded quality factor $Q_0$ of $\sim$~2700, was over-coupled to the waveguide so that its loaded quality factor $Q_L$ was $\sim$~100.   A triple-stub-tuner (TST), a custom made device with three perpendicular lengths of waveguides shorted by actuator controlled, moveable (0~to~25~mm) chokes, was inserted into the waveguide circuit outside the cryostat.  By changing choke positions, the TST allowed the impedance of the transmission line to be matched to the cavity and the central frequency chosen.  By this method the cavity was tunable, across the frequency range $\Delta f \, = \, f_0 / Q_L \, \sim \, 100$~MHz, where $f_0 \, \sim \, 12.91$~GHz is the central frequency of the cavity, while achieving at each point a resonance condition with a $Q$ value close to $Q_0$.

An antenna was attached to the back of the cavity in an under-coupled arrangement so as not to disrupt the field.  It picked up a signal (pickup) relative to the field inside the cavity and carried it through co-axial (SMA) cables to an I/Q mixer (MITEQ IR0618L C2Q) which displayed a DC signal proportional to the in-phase (I) and quadrature (Q) component of the input signal to a DSO.  From the I and Q signals, the time evolution of the amplitude and phase of the magnetic field in the cavity can be extracted for diagnostic perposes.  A second I/Q mixer analysed a sample microwave signal (sample), which was a fraction of the main signal attenuated by 50~dB, and displayed the information on the same DSO.

Timing was controlled by a Stanford Research (DG535) signal generator, triggered by a signal pulse from the AD.  Such a device was used to pulse the TWTA, to gate the PMTs and fire each of the Nd:YAG pump lasers.  Although the entire microwave pulse was 20~$\mu$s long the time for which the microwave signal affected the atoms was determined by the duration $T$ between the first and second laser pulse, which was of the order of several hundred nanoseconds starting after the microwave was switched on and ending before it was switched off.

Prior to the beam time, a range of microwave frequencies were determined with various TST positions and recorded.  In the $\sim$~90~s period between each $\pbar$ arrival, the data acquisition system was computer controlled to (1) move the TST to the next set of positions and set the VNA to the corresponding cw microwave frequency, (2)~warm up the TWTA at a rate of 10~Hz for 12~s, (3)~wait for a trigger from the AD, (4)~fire the amplifier and then the two lasers, (5)~record data.

\section{Results}\label{sec:Results}

To achieve the highest degree of precision it is necessary to make a high statistic measurement under optimum conditions.  The line width is limited by the Fourier transform of the microwave pulse which is determined by the laser delay time $T$, where the frequency difference between the half intensity points $\Delta \nu$~=~0.799$/T$.  Ideally, to make the line width as narrow as possible, the maximum laser delay should be chosen.  However, if $T$ is increased indefinitely, long before all the metastable atoms annihilate, the asymmetry, induced by the first laser, is lost due to relaxations caused to inelastic collisions and refilling from higher states.  Once an asymmetry has been created, the particles begin to return to equilibrium through spin exchanging inelastic collisions.  The relaxation time $\tau_c$ has been calculated by Korenman and Yudin to lie between 160.5~ns and 325~ns for a target density of 250~mbar and a temperature of 6~K \cite{Kman}.

\begin{figure}[!h]
\subfloat[]{
\label{fig:subfig:mwonoff}
\includegraphics[scale=0.4]{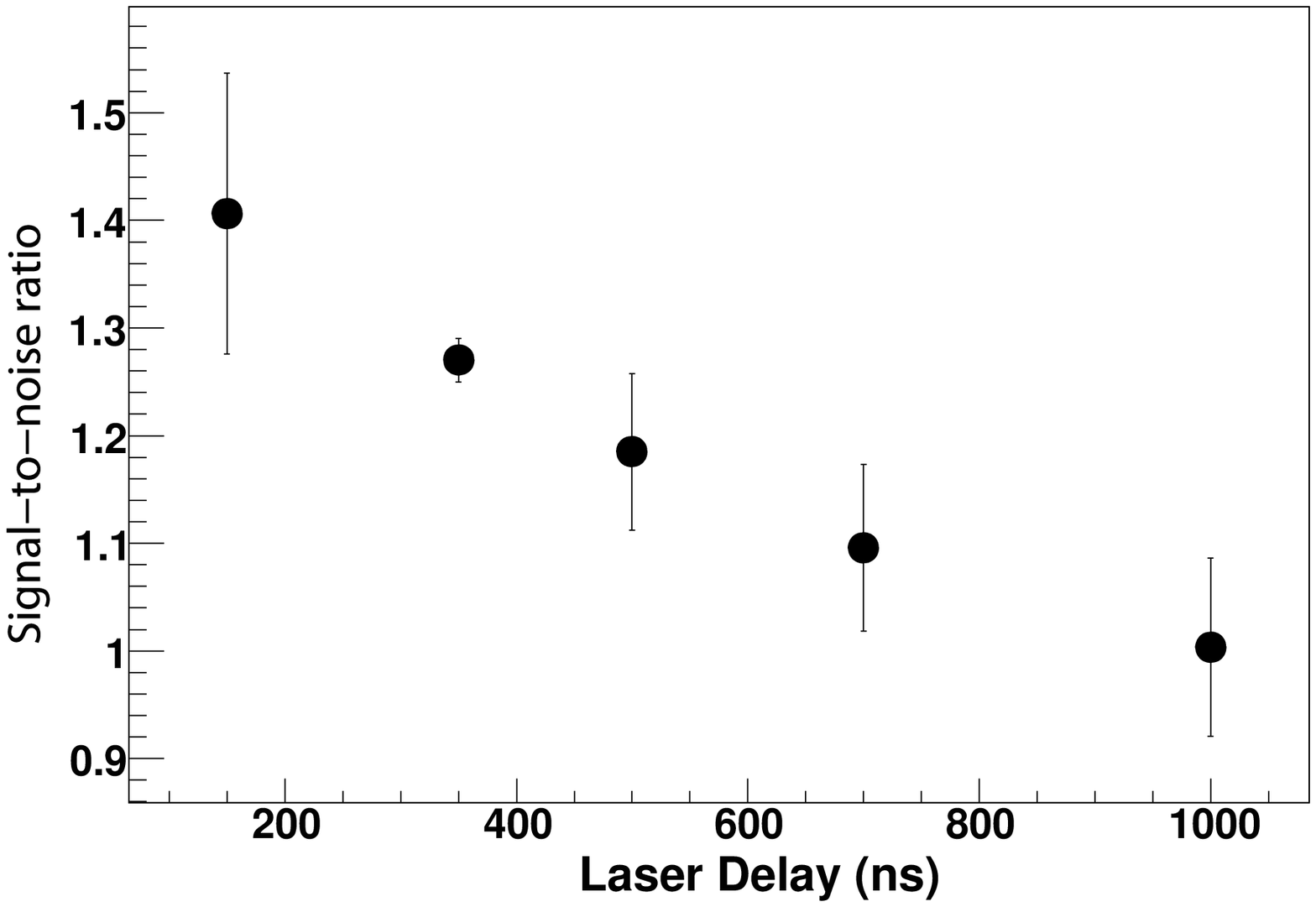}}
\subfloat[]{
\label{fig:subfig:width}
\includegraphics[scale=0.4]{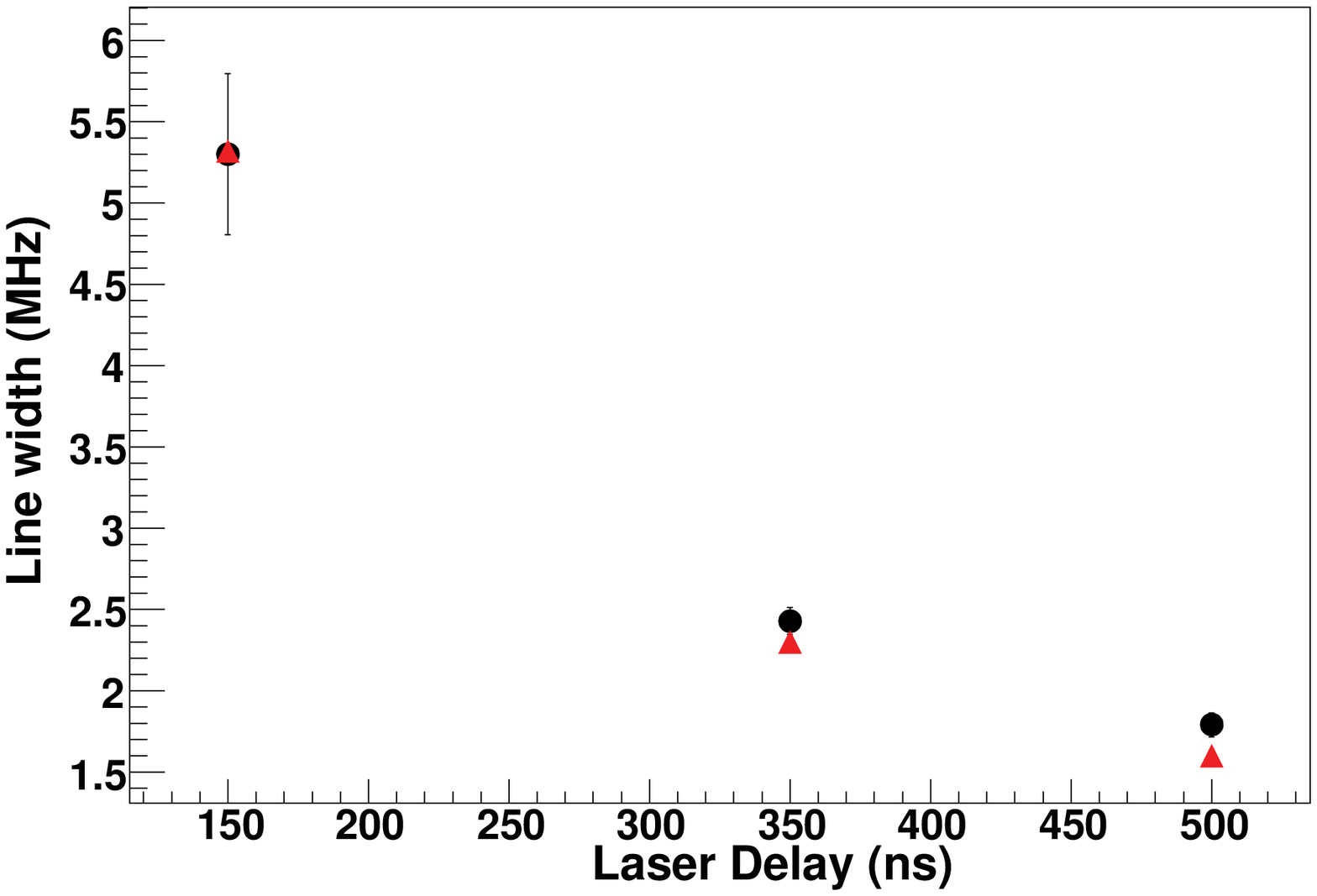}}

\caption{(a) The optimum signal-to-noise ratio measured for each time delay $T$, where the point at 350~ns has a reduced error because of higher statistics than the other points. (b) The line width as a function of $T$, where the widths of measured frequency scans are shown as circles and the widths calculated from the Fourier transform of the microwave time window as triangles.  Both sets of data were measured at a pressure of 250~mbar and a temperature of 6.1~K.}
\label{fig:MWinfo}
\end{figure}

To help determine the optimum conditions and to ensure that the asymmetry was not lost to collision induced relaxations, the maximum achievable signal-to-noise ratio was measured for a range of laser delays ($T$~=150~ns, 350~ns, 500~ns, 700~ns and 1000~ns).  The results are shown in figure \ref{fig:subfig:mwonoff} where the signal-to-noise ratio is defined as the ratio between the on resonance and off resonance signal amplitudes.  Scans at three different laser delays ($T$~=150~ns, 350~ns, 500~ns) were also measured, shown in figure \ref{fig:subfig:width}, where the reduction in line width can be observed with the increase in $T$.  In the $T$~=~350~ns and $T$~=~500~ns measurements the widths are just larger than the Fourier transform of the equivalent time delay. This small broadening is probably due to elastic collisions, calculated to be between  5.92~MHz and 2.66~MHz \cite{Kman}.

The microwave was scanned across a frequency range from 12.86~GHz to 12.94~GHz with a power of 4~W.  This power was carefully chosen so as to induce a $\pi-$pulse (half a Rabi oscillation) thus maximising the population inversion~\cite{Thomas}.  More points were measured over the two 15~MHz regions of each peak.  Less points were measured outside this region and were only used to confirm a flat background and determine the background level.  Each scan took approximately six hours and consisted of $\sim$~60 points, most of which were repeated three times.  One or two on-resonance microwave ``off'' points, where the microwave was fired at negligible power, were included in most of the scans to confirm the position of the baseline.

A total of 12~scans were measured with laser fluence of 30~mJ/cm$^{2}$ and $T$~=~350~ns at target density of 250~mbar and temperature of 6.1~K.  One scan was measured at $T$~=~500~ns at the same density and one scan was measured at $T$~=~350~ns at a target density of 150~mbar.  To reveal systematic errors the frequency points were measured consecutively and each $\pbar$ shot was monitored so that it could be remeasured if any component of the apparatus failed.  In addition to recording the ADATS, the data acquisition program logged other experimental variables, including the $\pbar$ beam profile, laser intensity and wavelength, microwave power and frequency.

The majority of systematic errors were caused by fluctuations of the $\pbar$ beam from the AD.  The BPM showed a time dependent drift away from the centre of the target which would change in direction and magnitude from shot to shot.  This resulted in a change of the signal, which was adjusted for by normalisation with the first laser and rejecting points when the $\pbar$ beam position and intensity fluctuated too greatly.  The laser wavelength was monitored during the measurement and continually adjusted to compensate for any drift.  The laser dye was changed regularly to ensure the power did not diminish with time.

\begin{figure}[!h]
\subfloat[]{
\label{fig:subfig:pickup}
\includegraphics[scale=0.4]{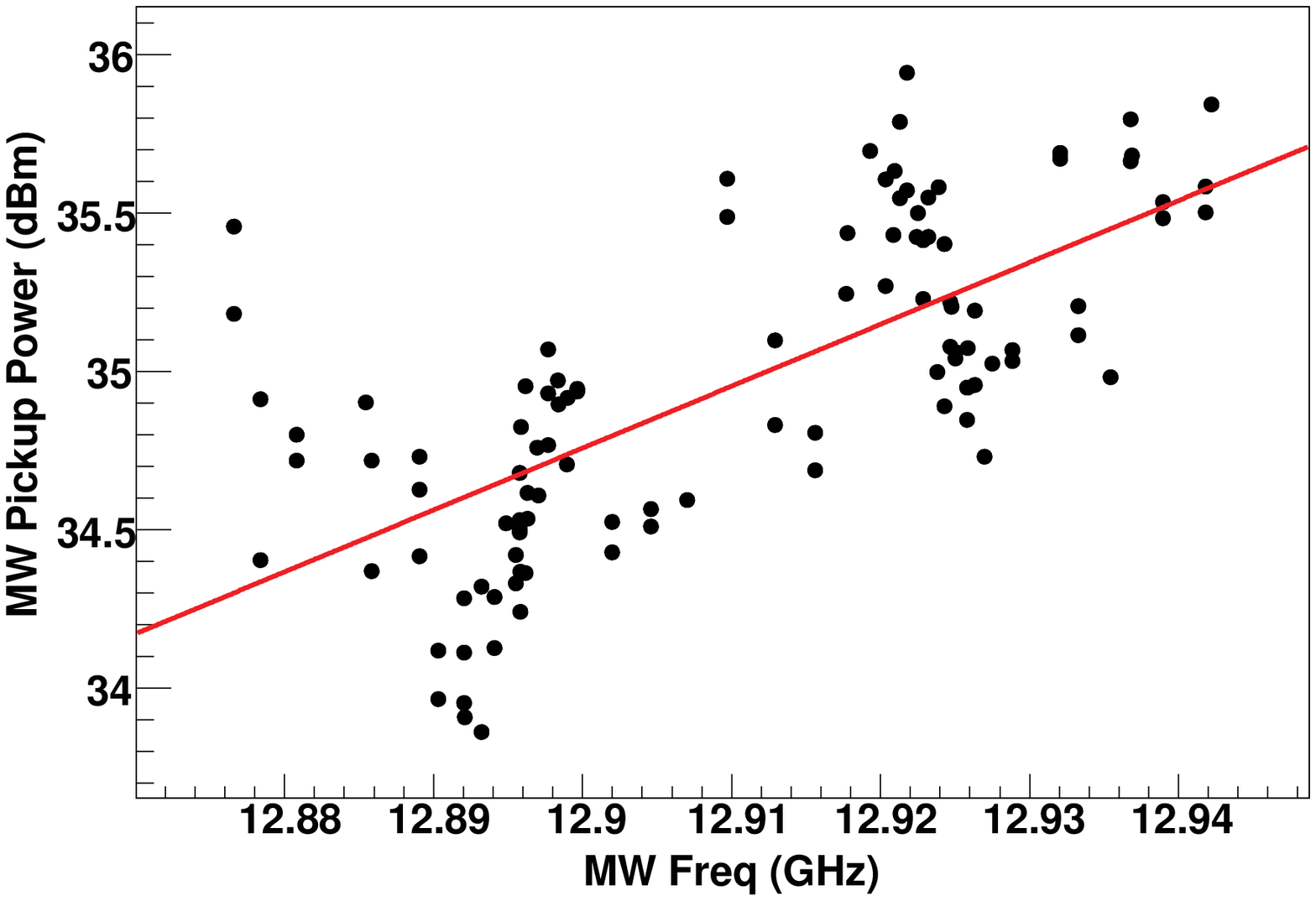}}
\subfloat[]{
\label{fig:subfig:sample}
\includegraphics[scale=0.4]{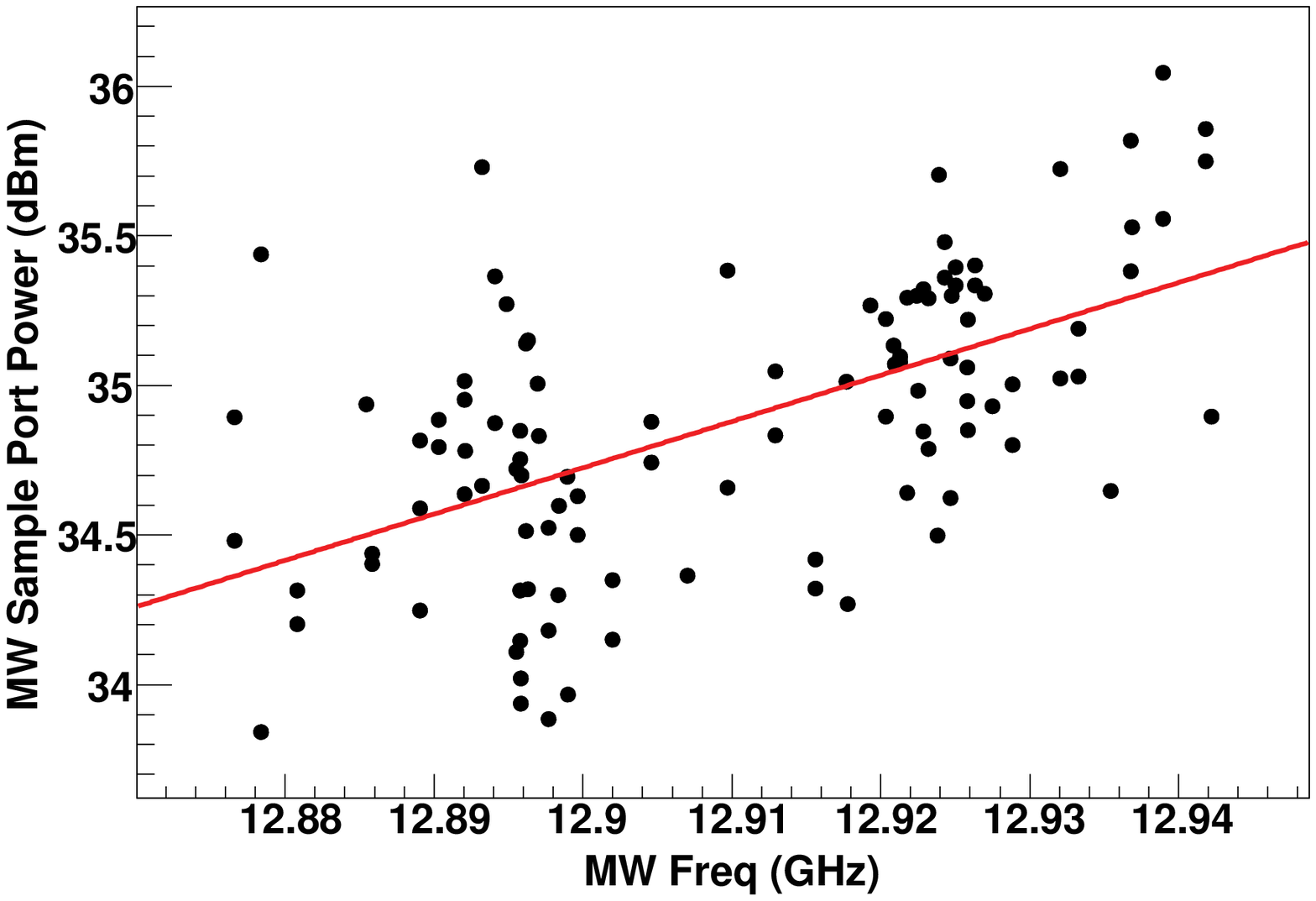}}
\caption{Microwave power vs frequency measured at two points: (a) antenna
pickup power after subtracting the frequency dependent coupling; (b)
microwave power at the sample port, adjusted for attenuation.  (0~dBm~=~1~mW).}
\label{fig:Power}
\end{figure}

The TWTA sample port power was observed to statistically fluctuate by $\sim$~10\% but also increased by an average of 1~dB (8\%) over the scanned region (figure~\ref{fig:subfig:sample}).  The same behaviour was observed on the pickup power at the antenna (figure \ref{fig:subfig:pickup}).  The change in power over the width of each peak was therefore considered to be small ($\sim$~20~W/GHz).  A simulation showed that the frequency dependence of the microwave power changed the fitted central frequency of the transition significantly only if the slope exceeded 300~W/GHz. 

The line shape for a two level system under the effects of an external oscillating field for time $T$ was expected to follow (\ref{eq:linewidth}).  The optimum case is when the system undergoes a $\pi$-pulse, $\mid b \mid T$ = $\pi/2$. Parameter $X(\omega)$ is the probability of transferring an atom to the other HF state and $b$ is related to the half-width of the distribution:
\begin{equation}\label{eq:linewidth}
X(\omega) = \frac{|2b|^2}{|2b|^2+(\omega_0-\omega)^2}\sin^2 \bigg\{ \frac{1}{2} \Big[|2b|^2+(\omega_0-\omega)^2 \Big]^{\frac{1}{2}}T \bigg\}.
\end{equation}
To confirm this was the correct shape, crosschecks of other functions were performed.  As the power had been measured to vary slightly between each resonant peak, two independent functions, rather than two identical ones, were used.  A Voigt function with a constant background was first used to determine the contributions from Gaussian, which is a good approximation of (\ref{eq:linewidth}), and Lorentzian functions.  The Lorentzian contribution was determined to be three orders of magnitude smaller than the Gaussian.

\begin{figure}[!h]
\begin{center}
\includegraphics[scale=0.5]{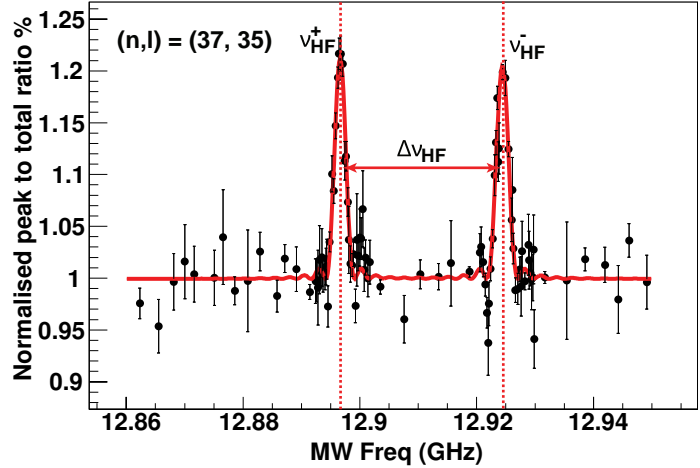}
\end{center}
\caption{Microwave resonance profile of the (37,~35) metastable state measured at a pressure of 250~mbar, a temperature of 6.1~K and a laser delay $T$~=~350~ns.  The $\nu^{+}_{\rm{HF}}$ and $\nu^{-}_{\rm{HF}}$ peaks are fitted with two independent equations of type (\ref{eq:linewidth}) and a constant background.} \label{fig:mwscan}
\end{figure}

\begin{figure}[!h]
\subfloat[]{
\label{fig:subfig:HF+}
\includegraphics[scale=0.4]{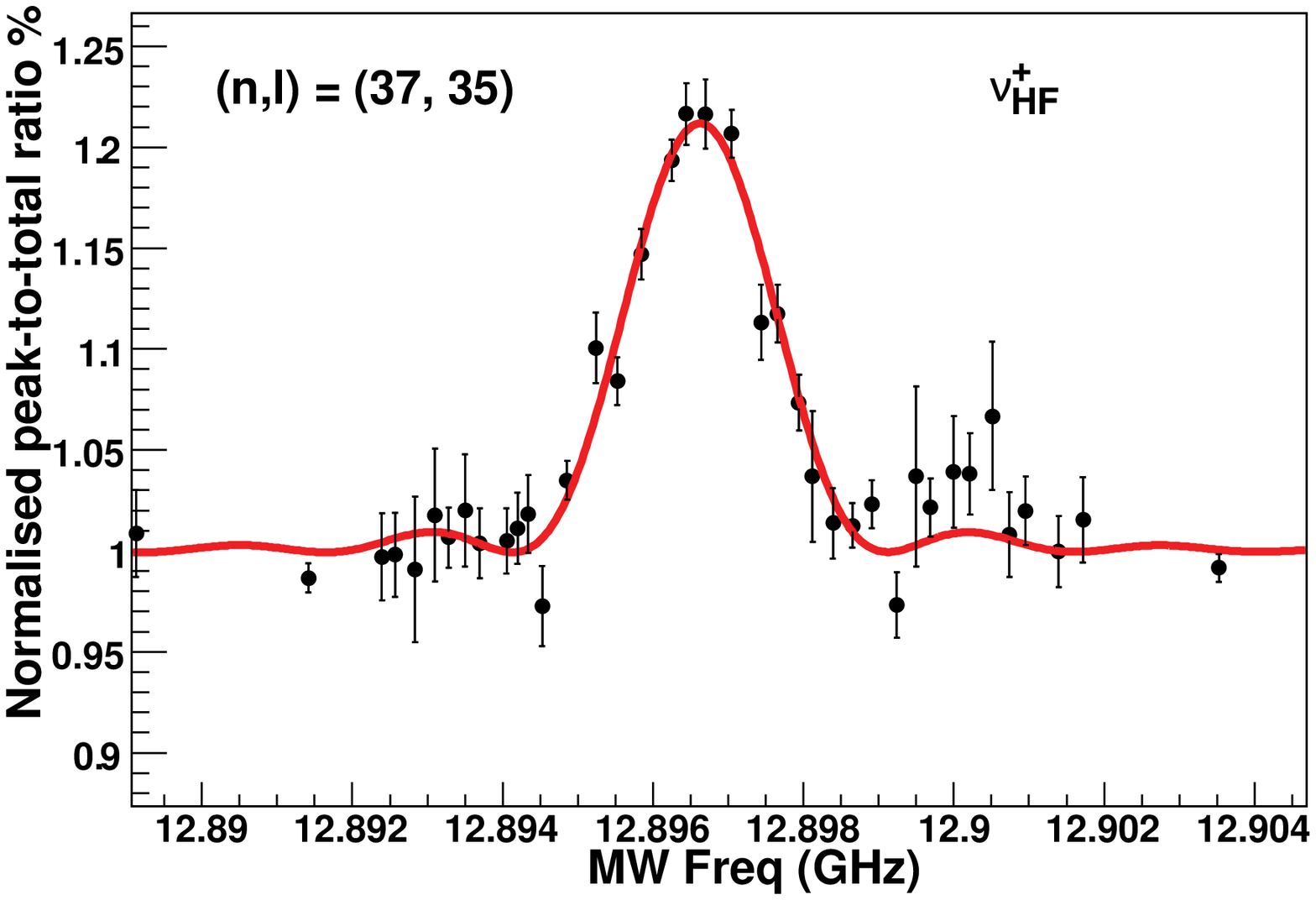}}
\subfloat[]{
\label{fig:subfig:HF-}
\includegraphics[scale=0.4]{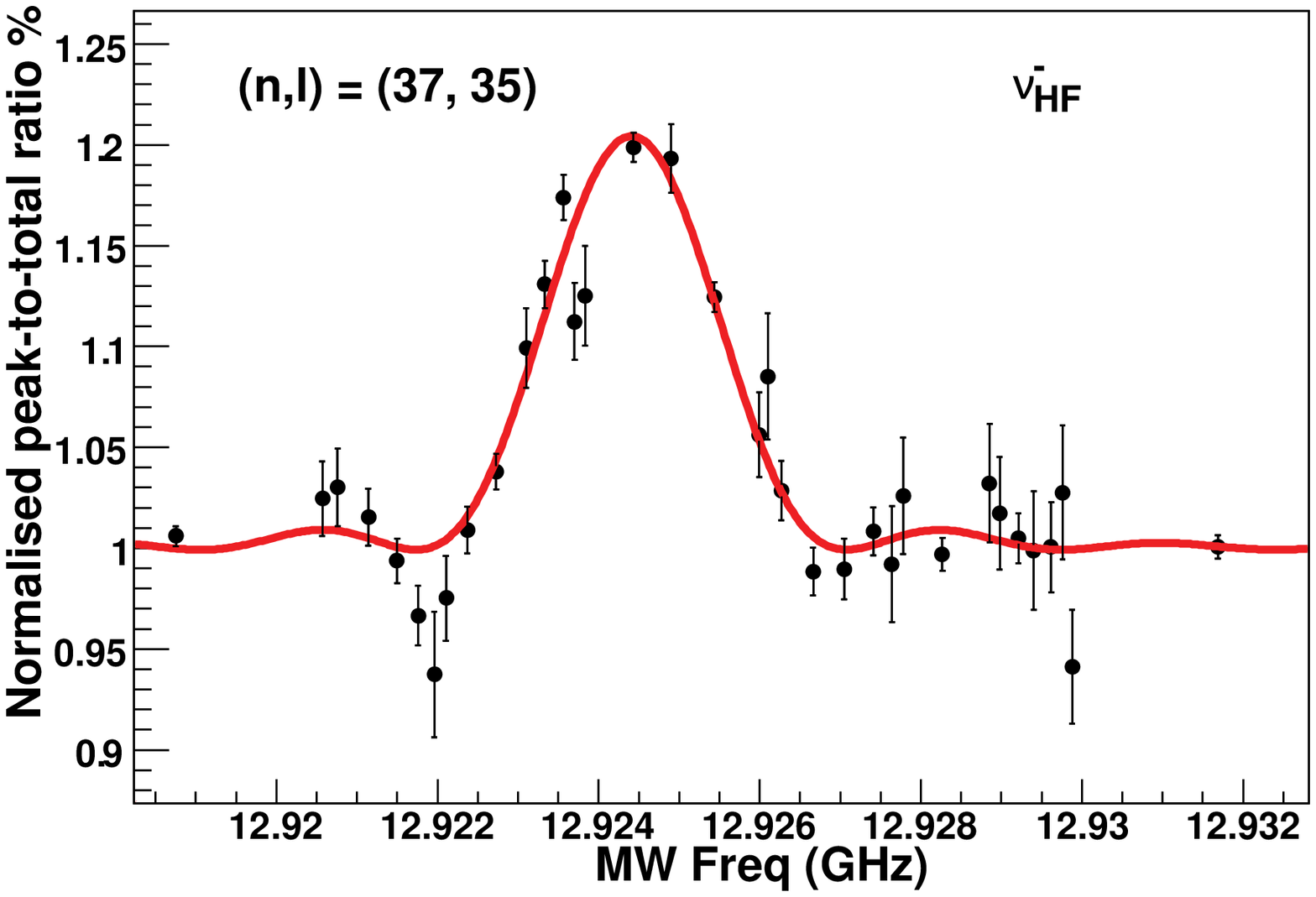}}

\caption{A close up of the HF peaks fitted with two independent equations of type (\ref{eq:linewidth}) plus a constant background, (a) $\nu^{+}_{\rm{HF}}$ and (b) $\nu^{-}_{\rm{HF}}$}
\label{fig:zoom}
\end{figure}

\begin{figure}[!h]
\begin{center}
\includegraphics[scale=0.5]{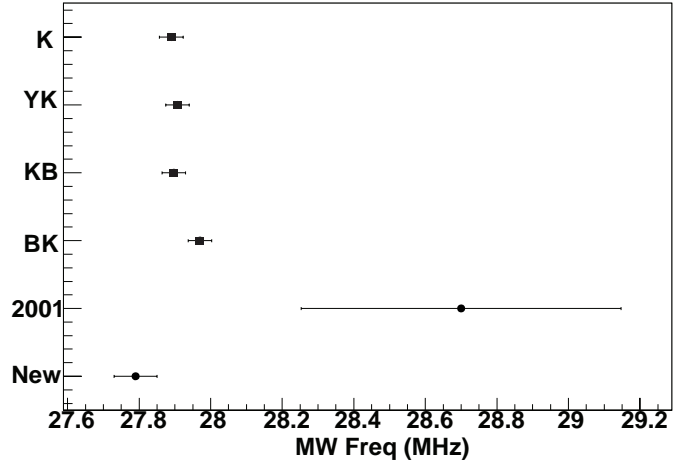}
\end{center}
\caption{Hyperfine splitting, comparing the recent result with the current theories (BK~\cite{BK}, KB~\cite{KB}, YK~\cite{YK},  K~\cite{K}) and our previous experiment (2001).  The error bars for theory are obtained from Bakalov and Widmann~\cite{BW}} \label{fig:splitting}
\end{figure}

The background was not assumed to be flat and was determined by fitting a sloped function.  This showed that the background increased slightly towards higher frequencies but the peak centres remained unchanged.  The slope of this background corresponds to the slope of microwave power shown in figure~\ref{fig:subfig:sample}.  However, as no Rabi oscillations should be induced off-resonance no effect of the microwave power should be observed outside the peaks and the observation is most likely to be statistical.
\begin{table}[!h]
\caption{Different fit results for the microwave scans, where the functions shown have been fitted as two independent functions with a constant background except ``Slope'' where two independent functions of type (\ref{eq:linewidth}) were fitted with a slanted background. Deg.~F is the number of degrees of freedom, Red.~$\chi^2$ is the reduced $\chi^2$ for the fits, $\nu^{\pm}_{\rm{HF}}$ are the HF transition frequencies,  $\Delta \nu_{\rm{HF}}$ is the difference between  $\nu^{+}_{\rm{HF}}$ and $\nu^{-}_{\rm{HF}}$.}\label{tab:fits}
\begin{indented}
\item[] \begin{tabular}{l  c c c  c  c  c }
\br
& $\chi^2$ & Deg.~F & red. $\chi^2$ & $\nu^{+}_{\rm{HF}}$ (GHz) & $\nu^{-}_{\rm{HF}}$ (GHz)  & $\Delta \nu_{\rm{HF}}$ (MHz)\\
\mr 
Voigt & 389.0 & 88 & 4.4 & 12.896 628(20) & 12.924 413(20) & 27.784(28) \\
Gauss & 388.8 & 89 & 4.4 & 12.896 628(20) & 12.924 412(20) & 27.785(28) \\
Eq.~(\ref{eq:linewidth}) & 349.4 & 89 & 3.9 & 12.896 622(20) & 12.924 412(19) & 27.789(28) \\
Slope & 348.0 & 88 & 4.0 & 12.896 623(20) & 12.924 412(20) & 27.788(28) \\
\br
\end{tabular}
\end{indented}
\end{table}
The different fitting methods all gave a similar result (see table~\ref{tab:fits}) but the two functions of type (\ref{eq:linewidth}) with a constant background fitted with the best $\chi^2$.  Figure~\ref{fig:mwscan} shows the averaged results of the 12~scans at target density of 250~mbar.  The fit shown is a function consisting of two independent equations of type (\ref{eq:linewidth}) with a constant background.  Figure~\ref{fig:zoom} shows a close up of each peak.  The results, with a numerical comparison to the current theories and previous experiment, are shown in table~\ref{tab:results} and a graphical representation of the HF splitting $\Delta \nu_{\rm{HF}} \, = \, \nu^{-}_{\rm{HF}} \, - \, \nu^{+}_{\rm{HF}}$ is displayed in figure~\ref{fig:splitting} where the error has been increased by the square root of the reduced $\chi^2$.


\begin{table}[h!]
\caption{Microwave hyperfine structure: experiment vs.~theory, where $\nu^{\pm}_{\rm{HF}}$} are the HF transition frequencies, $\Delta \nu_{\rm{HF}}$ is the difference between  $\nu^{+}_{\rm{HF}}$ and $\nu^{-}_{\rm{HF}}$, $\delta_{\rm{exp}}$ is the relative experimental error and $\Delta_{\rm{th-exp}}$ is the difference between theory and experiment.\label{tab:results}
\begin{tabular}{l  l  c  l  c  l  c}
\br
 & $\nu^{+}_{\rm{HF}}$ & $\delta_{\rm{exp}}$ & $\nu^{-}_{\rm{HF}}$ & $\delta_{\rm{exp}}$ & $\Delta \nu_{\rm{HF}}$ & $\delta_{\rm{exp}}$ \\

& (GHz) & (ppm) & (GHz) & (ppm) & (MHz) & ($\permil$)\\
\mr
Expt. & 12.896 622(39) & 3.0 & 12.924 412(39) & 3.0 & 27.790(55) & 1.9 \\
\mr
Expt.$_{2001}$ & 12.895 96(34) & 27 & 12.924 67(29) & 23 & 28.71(44) & 15.3 \\
\mr
 & & $\Delta_{\rm{th-exp}}$ & & $\Delta_{\rm{th-exp}}$ & & $\Delta_{\rm{th-exp}}$ \\
 & & (ppm) & & (ppm) & & ($\permil$) \\
\mr

BK~\cite{BK}  & 12.895 97 & -50 & 12.923 94 & -36 & 27.97 & 6.4\\
KB~\cite{KB} & 12.896 346 2 & -20 & 12.924 242 8 & -12.6 & 27.896 6 & 3.8\\
YK~\cite{YK} & 12.898 977 & 183 & 12.926 884 & 191 & 27.907 & 4.2\\
K~\cite{K} & 12.896 073 91 & -42 & 12.923 963 79 & -34 & 27.889 88 & 3.4\\
\br 

\end{tabular}
\end{table}

\section{Conclusions}\label{sec:Con}

This experiment initiates a systematic study of the HF splitting of the (37,~35) state of $\pbHe$.  Improvements to the laser system have increased the peak-to-total signal and reduced the line width  of the HFS lines to 2.4~MHz.  High statistics with stable conditions have reduced the experimental error by factor of five.  Although the new experimental value of the HF splitting is now smaller than its theoretical value, a convergence is observed.

It is predicted that the transition frequencies should shift by 80~kHz per 250~mbar \cite{Kman}.  The previous experiment, which had a resolution of 300~kHz, showed no evidence of a density shift \cite{HFShift} but, given the present resolution, this shift should be observable.  Due to time constraints, only preliminary measurements at 150~mbar were recorded and the statistics were insufficient to confirm or refute the expected shift.  However, as both HF lines shift equally, the upper limit of the density shift in the splitting $\Delta \nu_{\rm{HF}}$ is expected to be 7~kHz~\cite{Kman_priv}.  This effect is too small to be observed given that the experimental error is almost a factor ten greater.  To verify the magnitude of the predicted shift, a density and power dependent study is planned for 2008.

A comparison between the measured transition frequencies and three body QED calculations can be used to determine the antiproton spin magnetic moment.  The experimental precision required to reach the same accuracy as theory, for the (37,~35) state, is 33~kHz~\cite{BW}, a factor of two better than our current resolution.  A measurement to this degree of accuracy, if it agreed with theory, would provide the most precise determination of the antiproton to proton spin magnetic moment ratio to date (0.1\%). Other states offer even higher improvement factors, which are possible to measure with some slight changes to the experimental set up, and are planned for future years.

\ack

We are grateful to three undergraduate students: Andrea Pulm, Martin Leitgab and Gilbert Hangel who worked on this project.  We thank the AD operators for providing the antiproton beam.  This work was supported by Monbukagakusho (grant no. 15002005), by the Hungarian National Research Foundation (NK67974 and K72172) and by the Austrian Federal Ministry of Science and Research.

\end{document}